\documentstyle[12pt]{article}
\begin{document}

\newcommand{\bq}{\begin{equation}}
\newcommand{\eq}{\end{equation}}
\newcommand{\bqa}{\begin{eqnarray}}
\newcommand{\eqa}{\end{eqnarray}}
\newcommand{\nl}{\nonumber \\}
\newcommand{\eqn}[1]{Eq.(\ref{#1})}
\newcommand{\noi}{\noindent}
\newcommand{\dl}{\Delta}
\newcommand{\mapp}{M_{\mbox{{\small app}}}}
\newcommand{\sapp}{\sigma_{\mbox{{\small app}}}}
\newcommand{\umu}{^{\mu}}
\newcommand{\lmu}{_{\mu}}
\newcommand{\phigam}{\phi_{\gamma}}
\newcommand{\cgam}{c_{\gamma}}
\newcommand{\tmin}{t_{\mbox{{\small min}}}}
\newcommand{\tmax}{t_{\mbox{{\small max}}}}
\newcommand{\rlam}{\lambda(w^2,m^2,t)}
\newcommand{\eps}{\epsilon}
\newcommand{\vgam}{v_{\gamma}}
\newcommand{\plm}{$\pm$}

\pagestyle{empty}
\begin{flushright}
NIKHEF-H/94-01\\
CERN SL/94-03 (OP)\\
January 1994
\end{flushright}
\vspace{2cm}
\begin{center}\begin{Large}\begin{bf}
BBBREM -- Monte Carlo simulation \\
of radiative Bhabha scattering   \\
in the very forward direction
\end{bf}\end{Large}\\
\vspace{\baselineskip}
{\bf R.~Kleiss}\\
\begin{it} NIKHEF-H,
 P.O. Box 41882, 1009 DB Amsterdam, The Netherlands \end{it}\\
\vspace{\baselineskip}
{\bf  H.~Burkhardt}\\
\begin{it} SL Division, CERN,
  1211 Geneva 23, Switzerland \end{it}\\
\vspace{\baselineskip}
{\bf Abstract}
\end{center}
A fast and simple Monte Carlo program is presented that simulates
single Bremsstrahlung in Bhabha scattering,
$e^+e^-\to e^+e^-\gamma$, without constraints on scattering
angles. This allows the study of this process at arbitrarily
small, or even vanishing, scattering angles. Experimental cuts
can be imposed on an event-by-event basis, allowing for detailed
studies of the process as a limitation to beam lifetimes,
or a luminosity-measuring device, in $e^+e^-$ storage rings.
As an application, we show that the easy introduction of a
cutoff parameter, corresponding to the characteristic distance
between particles in the $e^{\pm}$ bunches, gives a reduced
cross section that is in good agreement with observation.
\begin{center}
{\it submitted to Computer Physics Communications}
\end{center}

\newpage
\pagestyle{plain}
\setcounter{page}{1}
\begin{center}
{\Large {\bf Program Summary}}
\end{center}

\noi{\it Title of program:\/} BBBREM\\


\noi{\it Program obtainable from:\/} R.~Kleiss,
 NIKHEF-H, P.O.~Box 41882, 1009 DB Amsterdam, The
 Netherlands, t30@nikhefh.nikhef.nl\\

\noi{\it Computers:\/}
systems supporting standard {\tt FORTRAN 77}\\

\noi{\it Programming language used:\/} {\tt FORTRAN 77}\\

\noi{\it Memory required:\/} about 200kb\\

\noi{\it number of bits per word:\/} 32\\

\noi{\it Subprograms used:\/} none\\

\noi{\it Number of lines in distributed program:\/} 386\\

\noi{\it Keywords:\/}
Bhabha scattering, Bremsstrahlung, radiative processes,
forward scattering, collinear singularities,
Monte Carlo simulation, experimental cuts,
beam lifetimes, luminosity monitoring\\

\noi{\it Nature of physical  problem:\/} Radiative Bhabha scattering,
$e^+e^-\to e^+e^-\gamma$, has a very large cross section at small
scattering angles, and plays various r\^{o}les in existing
and future $e^+e^-$ colliders. It can be an important background
to several two-photon scattering processes; it forms the major
ingredient in the finite lifetime of colliding beams; and it is
a possible process by which the luminosity can be measured, by
observing electrons or photons emerging at zero scattering angle.
Accurate knowledge of its cross section is therefore important.\\

\noi{\it Method of solution:\/}
Due to the extremely singular structure of the matrix elements,
and the possibility of complicated or unusual experimental cuts,
a straightforward integration of the cross section over the allowed
phase space is impractical. We therefore construct a Monte Carlo
algorithm that generates (random) events in phase space, with a
distribution that matches the actual cross section as closely as
possible. These events are assigned a {\em weight\/} which corrects
for discrepancies between the actual and the approximate matrix
elements: the average value of the weight in a sample of generated
events is the Monte Carlo estimate of the cross section.
Since each generated event is a complete description of the momenta
of the produced particles, any conceivable experimental cut can
be implemented, in addition to the single a-priori constraint,
namely, a minimum value for the energy of the Bremsstrahlung photon.
This value can be set (to essentially arbitrarily small values)
by the user. By setting
the weight of events that fail a particular
set of cuts to zero, one obtains
the cross section under those cuts.\\

\noi{\it Typical running time:\/} about 185 $\mu$sec per generated
  event on SUN SPARC 10;
  about 40 $\mu$sec per generated event on IBM 9000.\\

\noi{\it Unusual features of the program:\/} none

\vspace*{2cm}
\begin{center}
{\Large {\bf Long Write-Up}}
\end{center}
\section{Introduction}
The process of radiative Bhabha scattering,
\bq
e^-(p_1)\;e^+(q_1)\;\;\to\;\;e^-(p_2)\;e^+(q_2)\;\gamma(k)\;\;,
\label{theprocess}
\eq
where we have indicated the various momenta, is expected to play
an important r\^{o}le in several  physical problems. In the first
place, it is the main process by which electrons are lost upon the
collisions of $e^{\pm}$ beams, and dominates the beam lifetime
of LEP \cite{leplife}. Secondly, it may be an important background
to processes where single photons or $\pi^0$'s are observed
at small angles, such as at the DA$\Phi$NE collider
\cite{dafne}. One therefore intends to install tagging devices that
can, in principle, observe electrons and photons emitted at zero
scattering angle. Finally, this last device can also serve as a
luminosity monitor due to the very large event rate.
It is obvious that the various cross sections related to this process
have to be known accurately.

Various authors have already treated the process
(\ref{theprocess}), using a variety of approximations to
the matrix element. Since scattering over very small angles
completely dominates the process, Bremsstrahlung off the positron
and off the electron are essentially independent, and we use only
the two Feynman diagrams where the electron is seen to
radiate the photon. On the other hand, when all the
particles' trajectories are essentially collinear, the electron mass $m$
cannot be neglected with impunity. The result of the CALKUL collaboration
\cite{calkul}, where the leading correction terms of order $m^2/E^2$
are included (where $E$ is the beam energy), is also inapplicable
since it assumes that the electron mass is negligible with respect to
$t$, the momentum transfer of the positron.
Of the early publications, the work of Altarelli {\it et al.\/}
must be mentioned. They computed the total cross section
\cite{bucella} and various differential distributions \cite{stella}.
Similar results were obtained by a number of Russian authors
\cite{russians}. With the possibility of actually measuring the
process (\ref{theprocess}) at small or zero angle, these analytic,
totally inclusive, results are no longer adequate, and the
use of approximate matrix elements somewhat doubtful.
In \cite{matbab} a version of the matrix element, appropriate
to small-angle scattering, and correct up to truly negligible terms,
was described in detail.
In two recent publications, and Italian collaboration has presented
results for cross sections obeying a variety of angular and energy
cuts \cite{oresteplb}, and described the integration program
used to obtain these \cite{orestecpc} (the matrix element employed
in \cite{oresteplb} has, in fact, been checked to be completely
identical to that of \cite{matbab}).

In this paper we describe another approach to the study of
(\ref{theprocess}), namely that of a full-fledged Monte Carlo
simulation. The user generates any desired number of {\em events,\/}
that is, sets of momenta of the three produced particles. Their
distribution in phase space is matched as closely as possible to
the exact distribution. The remaining factor is included in
each event's {\em weight}. The average of the weights in the sample
is the Monte Carlo estimate of the cross section. This is a very flexible
approach since {\em any\/} experimental cut can be imposed by simply
setting the weight of events that fall outside the cuts to zero.
Moreover, since the momenta are explicitly given, the produced particles
can be tracked by other simulation programs, either into a detector,
or in the accelerator. A good check is provided by the fact that
those cuts that are provided for in \cite{orestecpc} must also yield
the cross section calculated by that program, by completely different
means\footnote{A small discrepancy was actually observed,
due to the fact that in \cite{orestecpc} the electromagnetic coupling
constant $\alpha$ used was 1/(138.318) instead of 1/(137.036) .}.

\section{The Monte Carlo}
We shall now describe our approach to the cross section
of \eqn{theprocess}. We start by defining the necessary variables.
We shall use the following invariants:
\bqa
& & s=(p_1\umu+q_1\umu)^2=4E^2\;\;\;,\;\;\;
s_1=(p_2\umu+q_2\umu)^2\;\;\;,\nl
& & t=(q_1\umu-q_2\umu)^2\;\;\;,\;\;\;
\dl_1=p_1\umu k\lmu\;\;\;,\;\;\;
\dl_2=p_2\umu k\lmu\;\;\;.
\eqa
Together with the azimuthal orientation of the event around the
beam axis, these serve to completely specify all the momenta.
In terms of these invariants, the correct form of the
matrix element, squared and summed/averaged over the particle spins,
is given by
\bqa
{1\over4}\sum\limits_{\mbox{{\small spins}}}|{\cal M}|^2
& = & e^6M\;\;,\nl
M  & = & {1\over\dl_1\dl_2|t|}
\left[s^2+s_1^2+(s+t-2\dl_2)^2+(s_1+t+2\dl_1)^2\right]\nl
& & -{4m^2\over\dl_1^2\dl_2^2}
\left[(s-2m^2)\dl_1-(s_1-2m^2)\dl_2-2\dl_1\dl_2\right]^2\nl
& & -{8m^2\over\dl_1\dl_2t^2}\left[\dl_1^2+\dl_2^2\right]\;\;,
\eqa
where $e$ is the electron charge.
The distribution of the generated events over the phase space shall
be proportional the {\em approximate} matrix element
\bq
\mapp = {4s^2C_1C_2\over\dl_1\dl_2|t|}\;\;,
\eq
where $C_1$ and $C_2$ are defined later on. This simple form
captures the essentials of the process dynamics, and also, owing
to clever definitions of the $C_{1,2}$, leads to a simple
result when integrated over the whole phase space. Denoting this
integral (or approximate total cross section) by $\sapp$,
we then define the weight  of a generated event by
\bq
\mbox{weight} = {M\over\mapp}\sapp\;\;,
\eq
for those events that satisfy the desired cuts, and zero otherwise.
Note here, that only one cut is always present, namely a lower limit
on the photon energy. This lower limit is given by
$\kappa E$, where $0<\kappa<1$, and serves to avoid the infrared
singularity lurking at $k^0=0$.\\

We shall deal with two Lorentz frames: the laboratory frame, and
the $R$-frame, that is, the rest frame of
$R\umu = p_2\umu+k\umu$. The most
convenient picture of the process (\ref{theprocess}) is then the
emission of a virtual photon by the positron (which is best
described in the lab frame), followed by `Compton' scattering
of the virtual photon and the electron (which finds its easiest
description in the $R$-frame). We therefore have
as `natural' invariants the following set:
\begin{itemize}
\item the components of the positron momentum transfer
  $q\umu=q_1\umu-q_2\umu$, namely $t=q\umu q\lmu$,
  $y=q^0/E$, and  $\phi$, the azimuthal angle of $\vec{q}_2$
  around the beam axis. These are all defined in the lab frame.
\item the scattering polar angle $\theta_{\gamma}$
  and azimuthal angle $\phigam$ of the photon
  with respect to the incoming electron momentum. These are defined
  in the $R$-frame, with $\cgam=\cos\theta_{\gamma}$.
\end{itemize}
In terms of these variables, the phase space integration
element can be written as
\bqa
d\Phi & \equiv & d^4p_2\delta(p_2^2-m_2)\;
  d^4q_2\delta(q_2^2-m^2)\;d^4k\delta(k^2)\;
  \delta^4(p_1+q_1-p_2-q_2-k)\nl
& = & {w^2-m^2\over32w^2}\;dy\;dt\;d\phi\;d\cgam\;d\phigam\;\;,
\eqa
where $w^2=R\umu R\lmu=sy+m^2$. The only nontrivial phase space bounds are
those on $t$: we have $\tmin \le t \le \tmax$, with
\bqa
\tmin & = & 2m^2-2Eq_2^0-2\left[
 (m^2-Eq_2^0)^2-m^2(E-q_2^0)^2\right]^{1/2}\;\;,\nl
\tmax & = & m^2sy^2/\tmin\;\;.
\eqa
Now, $y$ is typically very small, of order $m^2/E^2$; therefore,
$|\tmax|$ can become {\em extremely\/} small.

We may now write the invariants $\dl_{1,2}$ in terms of the chosen
variables:
\bqa
\dl_2 & = & sy/2\;\;,\nl
\dl_1 & = & sy\rlam(\eps+\vgam)/(4w^2)\;\;,\nl
\rlam & = & \left[(w^2+m^2-t)^2-4w^2m^2\right]^{1/2}\;\;,\nl
\eps & = & (w^2+m^2-t)/\rlam - 1\;\;,\nl
\vgam & = & 1-\cgam\;\;.
\eqa
Now, the approximate differential cross section is
\bq
d\sapp = {\alpha^3C_1C_2\over\pi^2|t|\rlam y(\eps+\vgam)}
 \;dy\;dt\;d\cgam\;d\phigam\;d\phi\;\;.
\eq
The angular integrals are now trivial, and we find
\bq
d\sapp = 4\alpha^3\log\left({2+\eps\over\eps}\right)
{C_1C_2\over|t|y\rlam}\;dy\;dt\;\;.
\eq
we now choose $C_1$ so as to simplify this expression:
\bq
C_1 \equiv \left[\log\left(1+{sy\over m^2}\right)\right]
           \left[\log\left({2+\eps\over\eps}\right)\right]^{-1}\;\;;
\eq
this approximation is justified when $|t|$ is very small. We
then have
\bq
d\sapp = 4\alpha^3\log\left(1+{sy\over m^2}\right)
{C_2\over|t|y\rlam}\;dy\;dt\;\;.
\label{tdist}
\eq
Now we do the $t$ integral; owing to our choice of $C_1$, it gives the
fairly simple result
\bqa
d\sapp & = & {4\alpha^3C_2\over sy^2}\log\left(1+{sy\over m^2}\right)
 \left[J(\tmax)-J(\tmin)\right]\;dy\;\;,\nl
J(t) & = & \log\left({\rlam+sy-t\over\rlam-sy-t}\right)\;\;.
\eqa
It is now time to choose $C_2$:
\bq
C_2 = 2\log\left({s\over m^2}\right)
      \left[J(\tmax)-J(\tmin)\right]^{-1}\;\;,
\eq
which goes to 1 in the typical case where $y$ is small.
The final integral is best performed in terms of the variable
$z\equiv sy/m^2$:
\bq
d\sapp = {8\alpha^3\over m^2}
 \log\left({s\over m^2}\right){\log(1+z)\over z^2}\;\;,
\label{zdist}
\eq
and, taking the upper limit on $z$ to infinity for simplicity,
we finally have
\bq
\sapp = {8\alpha^3\over m^2}\log\left({s\over m^2}\right)
\left[\log\left({1+z_0\over z_0}\right)
  -{\log(1+z_0)\over z_0}\right]\;\;,
\eq
where $z_0$ is the lower limit on $z$.
In practice, we want to impose a lower limit on $k^0$ rather
than on $z$, or $y$, since $q_2^0$ will usually be extremely close to
its maximum. However, for fixed $k^0$, the value of $y$ is bounded
by
\bq
y \ge {m^2\over s}\left({k^0\over E-k^0}\right)
\left({2\over1+\sqrt{1-{m^2\over E(E-k^0)}}}\right)\;\;,
\eq
which is reached when $\vec{p_2}$ and $\vec{k}$ are collinear and
opposite to $\vec{q_2}$. So, unless $\kappa$ is very close to $1$, we
have a lower bound
\bq
z \ge z_0 \equiv {\kappa\over1-\kappa}\;\;,
\eq
and we shall use this in the event generator. Note, however, that
upon generating events with this lower bound, we unavoidably also
get events with $k^0<\kappa E$. These we discard by putting their weight
to zero: even without additional experimental cuts, therefore, the
generated event sample will contain a modest fraction of events with
zero weight.\\

We are now in a position to generate the events. This is done by
applying mappings in the order opposite to that of the integration,
namely, first $z$, distributed according to \eqn{zdist}, then
$t$, according to \eqn{tdist}, and finally the angular variables,
of which only $\vgam$ is not completely trivial.
One minor remark is in order here: the generation algorithm assumes
an infinite upper limit on $z$. Very occasionally, therefore,
a $y$ value will occur that is unphysically large. This is
immediately indicated by $\tmin$ becoming complex, in which case the
event weight is put to zero and the event skipped. After having obtained
all the phase space variables, we can construct the momenta of the outgoing
particles. This is in principle straightforward, but is made tricky
by the enormous Lorentz boost necessary to go from the lab frame to the
$R$-frame and back. An ample collection of numerical pitfalls occurs,
which we avoid by using expressions in which the
major cancellations (from order 1 down to order $m^2/E^2$) have been
performed by hand. Finally, we calculate $M$ and $\mapp$, also
with an eye to the numerical stability. For the computation of $M$ in
an numerically stable manner, we refer to \cite{matbab}. In addition,
since $|t|$ is usually very small, $t^2$ will occasionally lead
to underflow problems. We therefore evaluate $t^2M$ and $t^2\mapp$
instead. Finally, we remark that ordinary double
precision ({\tt REAL*8}) suffices for our numerical work;
extended precision ({\tt REAL*16}), such as employed in
\cite{orestecpc}, is not necessary.

\section{Running {\tt BBBREM}}
The Monte Carlo has a very limited number of input parameters, read
at initialization from the standard input unit. They are, in order:
\begin{enumerate}
\item {\tt ROOTS} the total CMS energy of the incoming $e^+e^-$ system,
      in GeV. Any realistic value is allowed, from around 1 GeV (such as for
      the DA$\Phi$NE machine) up to hundreds of GeV for NLC/CLIC.
\item {\tt RK0} the photon energy cutoff fraction $\kappa$. In realistic
      cases, $\kappa$ will be of the order of percents, but it can be
      put much lower (although not to zero). When {\tt RK0} increases,
      the fraction of internally rejected events (see the discussion above)
      increases slightly. {\tt RK0=1} is not allowed.
\item {\tt NEVENT} the requested number of generated events. This includes
      events that come out with zero weight.
\item {\tt NRAN} a flag determining the source of (pseudo-)random numbers
      used. In principle, any reliable such source may be substituted
      instead of the routine {\tt RANDOM}, which occurs at 8 places
      in the generator code. We provide two alternatives:
      \begin{itemize}
      \item {\tt NRAN=1} A {\em very\/} crude, low-period
            multiplicative congruential algorithm. Its only merit
            is portability, and it should only be used for checking
            the program against the test run mentioned below.
      \item {\tt NRAN=2} A simple additive quasi-random number
            algorithm. We have used this to obtain the numerical results
            mentioned below. It is, however, not strictly portable
            since different rounding procedures on different machines
            lead to different sequences. The results below should, however,
            be reproduced to within statistical accuracy.
      \end{itemize}
\end{enumerate}
The input parameters {\tt ROOTS} and {\tt RK0} are transferred
to the event generator subroutine {\tt BCUBE} by common\\

{\tt COMMON/EXPERC/ ROOTS,RK0}\\

\noindent and {\tt NRAN} is transferred
to the random number source {\tt RANDOM} by common\\

{\tt COMMON/RANCHN/ NRAN}\\

\noindent When called the first time, {\tt BCUBE}
initializes, and prints the result for $\sapp$, in millibarn. Note
that this is {\em not\/} intended as an accurate numerical
approximation to the actual cross section! Upon each call, {\tt BCUBE}
fills the common\\

{\tt COMMON/LABMOM/ P1(4),P2(4),Q1(4),Q2(4),QK(4),D1,D2,T,WEIGHT}\\

\noindent which contains the essential information on the event:
the four-momenta $p_1\umu$, $p_2\umu$, $q_1\umu$, $q_2\umu$ and
$k\umu$ (the fourth component being the energy), all in GeV;
the invariants $\dl_1$, $\dl_2$, and $t$, all in GeV$^2$, whose
computation, directly from the returned momenta, is numerically
unstable; and, finally, the event weight, in millibarn. As given,
the main program yields, after the generation of {\tt NEVENT} events,
the following significant numbers:
\begin{enumerate}
\item {\tt RANCHK} the result of one last call to {\tt RANDOM}.
      This serves as a check on the portability
      of the program, since it depends both on the actual performance
      of the random number generator, and the {\em number\/} of
      random numbers that was actually used --- this number is variable
      due to various decision and rejection steps.
\item {\tt SMC} the computed cross section, {\it i.e.\/}
      the average weight.
\item {\tt SER} the estimated error on the value of {\tt SMC},
      defined in the standard way, using the variance of the weight
      distribution.
\end{enumerate}
In addition, some information on the weight distribution is given:
{\tt W1}, the sum of the weights; {\tt W2}, the sum of their squares;
{\tt WNIL}, the number of events with zero weight; {\tt WNEG},
the number of events with {\em negative\/} weight: these should not
occur, but if they do this would indicate rounding problems, or the
use of an inappropriate form for $M$ (for instance, negative weights
arise if the form of ref.\cite{stella} is used, although the cross
section is still essentially correct); and, finally, {\tt WMAX}, the
maximum weight occurring in the generation. Concerning this last
number it should be remarked that usually {\tt WMAX} is quite a bit
larger than {\tt SMC}: the weight distribution has a low tail.
This can be traced back to small values of $C_1$. In principle,
the $t$ integral in \eqn{tdist} can also be performed exactly, without
the benefit of $C_1$, leading to a slight complication in the algorithm.
This {\em might\/} be desirable in such cases as necessitate the
used of {\em unweighted\/} events, since high values of {\tt WMAX/SMC}
reduce the efficiency for unweighted events; for instance,
if the produced momenta are to be used for tracking in machine
studies. There, however, the tracking itself takes so much time that
a somewhat inefficient unweighted-event generation is not a problem.
We have therefore decided for the simpler algorithm. In any case, the
high-weight tail does not prohibit the attainment of a small Monte Carlo
error estimate.

\section{Results}
We shall now discuss some results obtained with {\tt BBBREM}.
In the first place, we reproduce here the output of a test
run with the following choice of parameters:
{\tt ROOTS=91.2d0}, {\tt RK0=0.01d0}, {\tt NEVENT=10000}, and
{\tt NRAN=1}. It is:\\

\begin{verbatim}
 ******************************************
 *** bbbrem : beam-beam bremsstrahlung  ***
 *** authors r. kleiss and h. burkhardt ***
 ******************************************
 total energy                    91.200000 gev
 minimum photon energy            0.010000 * beam energy
 approximate cross section    0.627848D+03 millibarn
 random: very crude random numbers chosen
 ******** cross section evaluation ********
 random number check       0.1471576690674
 total # of events                  10000.
 sum of weights               0.323102D+07
 sum of (weight**2)s          0.390034D+10
 max weight occurred          0.221525D+05
 # of zero weights                   1805.
 # of negative weights                  0.
 computed  cross section      0.323102D+03 millibarn
                   +/-        0.534452D+01 millibarn
 ******************************************
\end{verbatim}

\noindent We now turn to a more physical application of {\tt BBBREM},
concerning the lifetimes of $e^+e^-$ colliding beams. As discussed
in \cite{leplife}, the process (\ref{theprocess}) is actually
the dominant process by which electrons and positrons are lost from
the beams in storage rings such as LEP: an energy loss of
an electron/positron due to bremsstrahlung that
exceeds the so-called r.f. bucket half-height $s_b$
cannot be made up for by the r.f. system, and the particle will be
lost.  At LEP and most other $e^+e^-$ machines, $s_b$ is typically
about $10^{-2}$. One can therefore calculate, from first principles,
the corresponding cross section: at $\sqrt{s}=91.2$ GeV, corresponding
to LEP running at the $Z^0$ peak, one finds then $\sigma\sim 320$ mb,
both from {\tt BBBREM} and analytical calculations such as \cite{stella}.
Actual measurement \cite{leplife} finds, however, a cross section
of about 210 mb. As discussed in detail in \cite{leplife}, we
ascribe this reduction to a density effect: inside each bunch, the
electromagnetic fields from all electrons (say) overlap in such a
way that each electron has a finite interaction range, in contrast
to the usual infinite range. This range is of the order of half the
average distance $d$ between two electrons in the bunch (at rest!), which
is \cite{leplife} about 3.3 $\mu$m, corresponding to a momentum-transfer
squared $(\hbar c/d)^2 = 3.56\cdot10^{-21}$ Gev$^2$. Denoting this
number by $t_c$, we may then impose a cutoff on the momentum transfers
allowed in the process (\ref{theprocess}), in two ways:
\begin{itemize}
\item hard cutoff: events with $|t|<t_c$ are forbidden. In the Monte
  Carlo program this just means putting their weight to zero.
\item soft cutoff: the photon mediating the positron scattering
  is assigned an effective mass $\hbar c/d$, leading to an effective
  exponentially suppressed electromagnetic potential of the electrons.
  In the Monte Carlo, this is very simply implemented by multiplying
  the event weight with a factor $t^2/(t-t_c)^2$.
\end{itemize}
Note that it would be very hard to implement either of these two
cut-off procedures in any way but that of Monte Carlo.
Below we present results for four different values of {\tt ROOTS}:
1.02 GeV, corresponding to DA$\Phi$NE at the $\Phi$-meson peak,
60 GeV, appropriate to TRISTAN,
91.2 GeV, corresponding to LEP 1 at the $Z^0$ peak,
and 190 GeV, typical for LEP200.  In each case we have
used {\tt RK0=0.01}, {\tt NEVENT=1000000}, and {\tt NRAN=2}.
The cross sections are given in millibarn, for each of the
three cases: no cutoff, hard cutoff, soft cutoff.\\

\begin{tabular}{|l|c|c|c|}\hline\hline
{\tt ROOTS} (GeV) & no cutoff & hard cutoff & soft cutoff \\ \hline
1.02     &  208.1    &   198.7     &  194.6      \\
60       &  308.6    &   210.2     &  204.0      \\
91.2     &  318.9    &   210.1     &  203.9      \\
190      &  336.8    &   209.8     &  203.7      \\
\hline\hline\end{tabular}

\vspace*{\baselineskip}
\noindent The estimated error is 0.4 millibarn for DA$\Phi$NE, and 0.5
millibarn in the other cases.
We see that the LEP 1 cross sections with the cutoff are in much better
agreement with the observation. For more details about this
physical, rather than computational, issue we refer to \cite{leplife}.

\end{document}